\documentclass[twocolumn,aps,showpacs,preprintnumbers,amsmath,amssymb,superscriptaddress]{revtex4}

\usepackage{graphicx}
\usepackage{dcolumn}
\usepackage{bm}

\begin{document}

\title{Timescale for equilibration of N/Z gradients in dinuclear systems}

\author{K. Brown}
\affiliation{
Department of Chemistry and Center for Exploration of Energy and Matter\\
2401 Milo B. Sampson Lane, Bloomington IN 47405, USA}

\author{S. Hudan}
\email{shudan@indiana.edu}
\affiliation{
Department of Chemistry and Center for Exploration of Energy and Matter\\
2401 Milo B. Sampson Lane, Bloomington IN 47405, USA}

\author{R.~T. de Souza}
\email{desouza@indiana.edu}
\affiliation{
Department of Chemistry and Center for Exploration of Energy and Matter\\
2401 Milo B. Sampson Lane, Bloomington IN 47405, USA}

\author{J. Gauthier}
\affiliation{Universit{\'e} Laval, Qu{\'e}bec, Canada}

\author{R. Roy}
\affiliation{Universit{\'e} Laval, Qu{\'e}bec, Canada}

\author{D.~V. Shetty}
\affiliation{Cyclotron Institute, Texas A\&M University, College Station, TX 77843, USA}

\author{G.~A. Souliotis}
\affiliation{Cyclotron Institute, Texas A\&M University, College Station, TX 77843, USA}
\affiliation{Laboratory of Physical Chemistry, Department of Chemistry, 
National and Kapodistrian University of Athens, Athens 15771, Greece}

\author{S.~J. Yennello}
\affiliation{Cyclotron Institute, Texas A\&M University, College Station, TX 77843, USA}

\date{\today}
\begin{abstract}
Equilibration of N/Z in binary breakup of an excited and transiently deformed 
projectile-like 
fragment (PLF$^*$), produced in peripheral collisions 
of $^{64}$Zn + $^{27}$Al, $^{64}$Zn, $^{209}$Bi
at E/A = 45 MeV, is examined. The composition of emitted light fragments (3$\le$Z$\le$6)
changes with the decay angle of the PLF$^*$. The most neutron-rich fragments observed
are associated with a small rotation angle. A clear target dependence is observed with
the largest initial N/Z correlated with the heavy, neutron-rich target. Using the rotation 
angle as a clock, we deduce that N/Z equilibration persists for times as long as 
3-4 zs (1zs = 1 x 10$^{-21}$s = 300 fm/c). The rate of N/Z equilibration is found to 
depend on the initial neutron gradient within the PLF$^*$.

\end{abstract}


\maketitle

The broad impact of the density dependence of the nuclear symmetry energy
makes it a topic of considerable interest. 
Whether the asymmetry term 
in the nuclear equation of state follows a stiff or a soft dependence on density 
determines the composition of a neutron star's crust, and the conditions under which
a supernova explosion occurs \cite{Lattimer01,Steiner05,Janka07}.
In the case of nuclei, the stability of the heaviest elements \cite{Moller12}, 
and the existence of neutron skins \cite{Tsang12}, also depend on this quantity. 
One means of 
investigating the density dependence of the symmetry energy is by measuring the N/Z equilibration
in a dinuclear system \cite{Tsang04,Keksis10,Kohley12}. 
Although past studies have principally focused on the N/Z equilibration that occurs between
the projectile and target nuclei in a collision, such an approach is fundamentally 
limited by 
the short contact time between the two collision partners. This contact time, which is
approximately 100 fm/c at intermediate energies, inherently limits the degree of equilibration that 
can be attained.

Another opportunity to 
investigate N/Z equilibration is the
dynamical binary breakup of an excited and transiently deformed nucleus produced
in the semi-peripheral collision of two heavy-ions at intermediate energies 
(E/A=20-100 MeV) \cite{deFilippo12}. 
At these energies, the collision  
of two heavy-ions at 
peripheral and mid-central impact parameters leads to 
the exchange of charge, mass, and energy 
between the projectile and target nuclei.
Prior experimental and theoretical work demonstrates that several mechanisms can contribute to the
ternary breakup of the transient system into two heavy 
remnants (projectile and 
target)
together with one intermediate mass fragment \cite{Montoya94, Colin03, deFilippo12, Baran12}. For many such breakups, the existence 
of a short-lived neck joining the projectile and target plays a crucial role. As the two heavy remnants separate
the neck ruptures leaving it still attached to one of the two remnants. The binary system consisting 
of the 
neck and associated remnant continues to undergo neutron and proton exchange as it rotates.
If the neck is associated with the projectile remnant,
we designate the joint neck-heavy remnant system as the projectile-like fragment (PLF$^*$).
In the case of ternary decays a second rupture of the neck occurs producing the intermediate mass fragment.
Statistical decay from the equilibrated projectile and target remnants also occurs.

Equilibration of N/Z 
within the PLF$^*$ has been observed to  
persist for timescales as long as t $\approx$ 3zs (1zs = 1 x 10$^{-21}$s) \cite{Hudan12}.
In the present work we further characterize this approach, extending it to a significantly 
lighter projectile. We explore the impact of the N/Z of the target on the composition of 
the fragments 
produced and their subsequent evolution. We also demonstrate that this 
equilibration process exists when the projectile and target nuclei 
are the same nucleus.

The experiment was conducted at the Cyclotron Institute at Texas A\&M University, where 
a beam of $^{64}$Zn ions was accelerated to 
E/A = 45 MeV with an average beam intensity of $\approx$2x10$^{8}$ p/s. 
The beam
impinged on $^{27}$Al, $^{64}$Zn and $^{209}$Bi targets with thicknesses of 
13.4, 5, and 1 mg/cm$^2$ respectively.
Although
the experimental details have been previously published 
\cite{Theriault06}, they are summarized below for completeness.
The array
FIRST \cite{FIRST},which subtended the
angular range 4.5$^\circ$ $\le$ $\theta_{lab}$ $\le$ 27$^\circ$ was
used to identify
charged products produced in the reaction. 
In the angular range 
4.5$^\circ$ $\le$ $\theta_{lab}$ $\le$ 7$^\circ$, 
the forward telescope in FIRST
provided identification by atomic number of all products up to 
Z=30 and isotopic information for Z $\le$ 12. 
The second telescope in FIRST 
, which subtended the angular range
7$^\circ$  $\le$ $\theta_{lab}$ $\le$ 14$^\circ$, 
provided Z identification for Z $\le$ 22 and A identification for
Z $\le$ 8. The third telescope in FIRST 
(14$^\circ$  $\le$ $\theta_{lab}$ $\le$ 27$^\circ$)
provided Z identification for Z $\le$ 12 and A identification for
Z $\le$ 7.
The high segmentation of FIRST provided an 
angular resolution of 
$\pm$0.05$^\circ$(4.5$^\circ$  $\le$ $\theta_{lab}$ $\le$ 7$^\circ$),
$\pm$0.44$^\circ$(7$^\circ$  $\le$ $\theta_{lab}$ $\le$ 14$^\circ$) and 
$\pm$0.81$^\circ$(14$^\circ$  $\le$ $\theta_{lab}$ $\le$ 27$^\circ$) in polar angle
and $\pm$11.25$^\circ$ in azimuthal angle.
The energy resolution obtained was approximately 1\%.

In order to focus on binary decays,
events were selected in which two fragments (Z $\ge$ 3) were detected within 
the laboratory angular range
4.5$^\circ$ $\le$ $\theta_{lab}$ $\le$ 27$^\circ$. 
These two 
fragments were distinguished from each other by their atomic number, with the 
larger (smaller) atomic fragment designated as 
Z$_H$ (Z$_{L}$). 
We ensured that the PLF$^*$ under investigation 
comprised a large fraction of the initial projectile by requiring that
the 
events selected had Z$_{H}$ $>$ 11. 
Events selected in this manner corresponded to approximately 14$\%$ of the measured
yield in which one fragment with Z$>$11 was detected.

\begin{figure}
\includegraphics[scale=0.66]{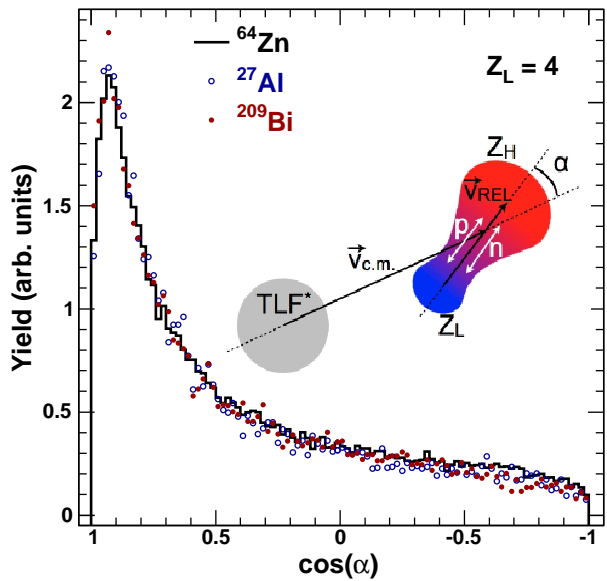}
\caption{\label{fig:cosalpha} (Color online) The angular distribution 
of binary splits (Z$_L$-Z$_H$)
for Z$_{L}$=4, representative of other fragments, 
is shown.
Data for the $^{64}$Zn, $^{27}$Al, and $^{209}$Bi targets
are represented by the black line, blue open symbol and red closed symbol
histograms respectively.
}
\end{figure}

It has previously been established 
that 
an instructive quantity for the binary decay of the PLF$^*$ is 
the angle between the 
direction of the two fragments center-of-mass velocity, v$_{c.m.}$, and their 
relative velocity, v$_{REL}$, defined as v$_{H}$-v$_{L}$ \cite{Davin02, McIntosh10}. 
We construct the 
angle $\alpha$, with

$cos(\alpha) = v_{c.m.}\cdot v_{REL}/(\|v_{c.m.}\|\|v_{REL}\|)$ 

as indicated within the inset of
Fig.~\ref{fig:cosalpha}.
Consequently, aligned decays 
with Z$_L$ emitted backward (forward) of Z$_H$ corresponds to
cos($\alpha$) = 1 (-1). 
Momentum correlations observed 
between Z$_H$ and Z$_{L}$ reveal that these
two fragments originate from a common parent as evident in Fig.~2 of \cite{McIntosh10}. 
This parent nuclear system comprised of
Z$_H$ and Z$_{L}$ is designated as the PLF$^*$. 

The angular distribution for Z$_L$=4 fragments is presented in 
Fig.~\ref{fig:cosalpha} for the $^{64}$Zn, $^{27}$Al, and $^{209}$Bi targets used.
Consistent with previous work, all the angular distributions manifest a peak at backward angles 
cos($\alpha$)$>$0.5. 
This preferential backward decay of the
PLF$^*$ is well established, and has been interpreted as the aligned dynamical 
decay of the excited and 
transiently deformed PLF$^*$ \cite{Glassel83, Lecolley95, Davin02}. 
The backward peaking of the angular distribution 
can be understood as the dynamical binary splitting of the PLF$^*$ on a timescale 
that is short relative to the rotational period of the PLF$^*$.
To compare the shape for the different targets,
all three distributions have been normalized in the 
interval -1$\le$cos($\alpha$)$<$0.
We chose this region for normalization since it corresponds to forward 
statistical emission from the PLF$^*$.
This forward statistical decay is long-lived relative to backward emission and hence 
is less coupled
to any dynamics responsible for the formation of the PLF$^*$.
For each target, the angular distributions exhibit the same shape, manifesting a 
distinct preference for aligned decay of the PLF$^*$
with the Z$_L$ fragment oriented towards the target. The shape of the
distribution for cos($\alpha$)$<$0 provides an indication that the angular momentum
of the decaying PLF$^*$ is relatively small. In contrast to previously studied 
systems \cite{Hudan12}, the yield does not increase near cos($\alpha$)=-1.

Given the normalization at forward angles, the similarity of the distributions 
for the three targets at backward angles, cos($\alpha$)$>$0, is striking.  
This similarity suggests that while the probability of forming the elongated and excited PLF$^*$,
as well as its composition,  may depend on the target, the relative probability of its 
subsequent decay is essentially independent of the target.

\begin{figure}
\includegraphics[scale=1.0]{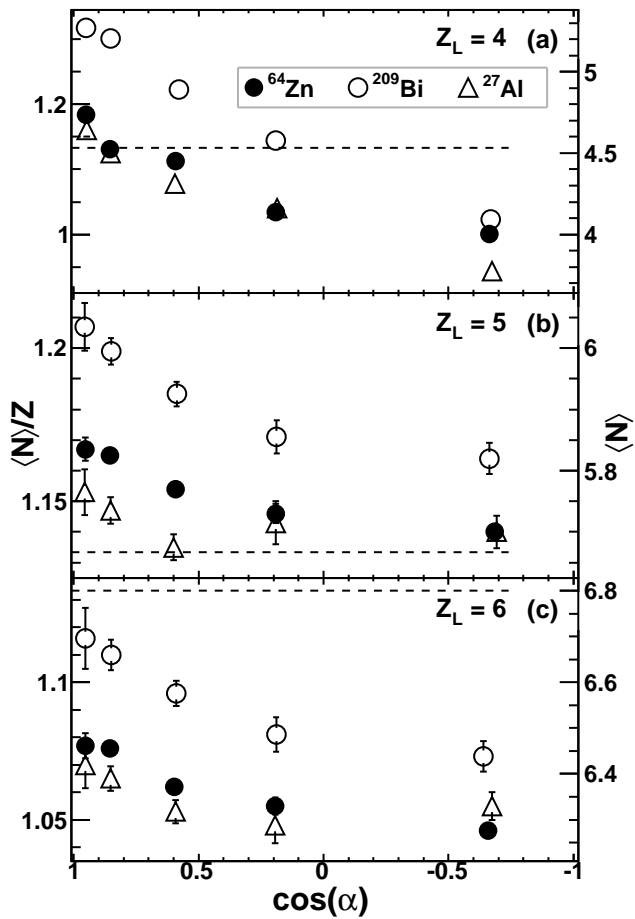}
\caption{\label{fig:NZ_CosAlpha}
Average neutron to proton ratio for selected Z$_{L}$ as a function of
the decay angle. The ratio for the $^{64}$Zn, $^{209}$Bi, and $^{27}$Al targets
is represented by the closed circle, open circle and open triangle respectively.
The dashed line for each Z$_{L}$ represents the N/Z of the 
$^{64}$Zn beam.
}
\end{figure}

We next examine whether the composition of the Z$_L$ fragment changes 
as a function of rotation angle.
In our initial work, which analyzed the reaction 
$^{124}$Xe + $^{112,124}$Sn,
we observed that the
$\langle$N$\rangle$/Z of the Z$_L$ fragment
decreased as the Z$_L$-Z$_H$ system rotated \cite{Hudan12}. 
As the Zn-like PLF$^*$ in the present work 
is considerably
smaller than the Xe-like PLF$^*$, it was unclear whether 
the behavior previously observed would also 
exist for the smaller PLF$^*$. Depicted in 
Fig.~\ref{fig:NZ_CosAlpha} is the $\langle$N$\rangle$/Z of the Z$_L$ 
fragment for Be (panel a),
B (panel b), and C (panel c) fragments. For each Z$_L$ shown the impact of 
the three different targets
is also presented. 
The average $\langle$N$\rangle$/Z for a given Z$_L$ is deduced by averaging 
the neutron number for the different isotopes measured.
A common feature of all the data is that the largest value 
of $\langle$N$\rangle$/Z is associated with cos($\alpha$)=1, namely backward emission. 
As the Z$_L$-Z$_H$ system rotates, $\langle$N$\rangle$/Z of the Z$_L$ 
fragment decreases 
corresponding to a net loss of neutrons by the Z$_L$ 
fragment. 
In the case of the Be fragments, this dependence of 
$\langle$N$\rangle$/Z on cos($\alpha$) is clearly apparent 
even for the lightest 
target, Al. For all three fragments shown
the magnitude of $\langle$N$\rangle$/Z is largest for the Bi target. 
We attribute this large value of $\langle$N$\rangle$/Z for 
cos($\alpha$)$\approx$1 in the case of the Bi target to the 
preferential pickup of neutrons by the PLF$^*$ from the Bi target with its N/Z=1.51. 
In contrast, the $^{64}$Zn and $^{27}$Al targets with N/Z = 1.13 and 1.07 do not 
present a neutron-rich reservoir from which the PLF$^*$ can
pick up neutrons.

Shown in the right hand scale of 
Fig.~\ref{fig:NZ_CosAlpha} is the $\langle$N$\rangle$ of the 
Z$_L$ fragment. In the case of the Be fragments, for the Bi target $\langle$N$\rangle$ 
decreases from 5.2 to 4.05 a net change of over one neutron. For the
Zn and Al targets, a somewhat smaller net decrease of 0.6 - 0.7 in neutron 
number is observed. For the Bi target, the change in $\langle$N$\rangle$ for
Z$_L$=5 and Z$_L$=6 is $\approx$0.2. 
The change in $\langle$N$\rangle$ for Li fragments and the Bi target 
(not shown) is also $\approx$0.2, comparable to that of B and C fragments. 
The larger change observed in the case of Be fragments can be qualitatively 
understood as being due to the absence of $^8$Be fragments. Since the isotopic 
distribution for all the fragments with Z$_L$=3,5, and 6 has a value of
$\langle$N$\rangle$/Z$>$1, it is reasonable to expect that this is also the 
case for Be fragments. The decay of $^8$Be into two alpha particles removes 
these fragments from the measured isotopic distribution thus artificially 
increases the value of $\langle$N$\rangle$ observed for Be at 
backward angles. 
In effect, the absence of $^8$Be acts as an amplifier for the change in 
$\langle$N$\rangle$ by emphasizing the importance of the extremes of the 
isotopic distribution. This conclusion is supported by our re-analysis
of carbon isotopes in which we eliminate $^{12}$C from the isotopic 
distribution\cite{Supplemental_NZ}.
For this reason, we have elected to 
present the Be data without correcting for the absence of $^8$Be.

The physical picture that emerges is one in which the N/Z of the dinuclear PLF$^*$
is established through its interaction with the target. Preferential 
transfer of neutrons from a 
neutron-rich target such as Bi results in a neutron-rich PLF$^*$. As the 
nascent Z$_L$ 
fragment is oriented towards the target-like fragment, it is the primary beneficiary of 
the transferred 
neutrons. 
In addition, even for a symmetric projectile-target collision, the density 
dependence of the symmetry energy leads to neutron enrichment of the low-density 
neck\cite{Theriault06}.
The result is an initial N/Z gradient within the PLF$^*$. 
As time passes, these additional neutrons in the
Z$_L$ fragment are dissipated. 
Whether this preferential neutron transport out of the 
Z$_L$ fragment 
occurs into the Z$_H$ fragment
or into a low-density neck region connecting the Z$_L$ and Z$_H$ fragments 
is presently unclear \cite{Baran12}. 
It should be clear that transfer of both neutrons and protons occurs between the
Z$_H$ and Z$_L$ fragments. Our selection of a particular Z$_L$ fragment 
in this analysis precludes 
us from examining the net proton exchange.

If the decrease in $\langle$N$\rangle$/Z with cos($\alpha$) can be understood 
as the preferential 
transport of neutrons out of the Z$_L$ fragment, one might expect
that the shorter the contact time between the Z$_L$ and Z$_H$ 
fragments the less likely it is that the initial $\langle$N$\rangle$/Z 
is decreased.
Dynamical splitting of the dinuclear Z$_H$-Z$_L$ system can be viewed as 
a dynamical fission process in which the
reaction dynamics provides collective motion along the separation axis of the Z$_L$-Z$_H$ 
system \cite{Davin02, Colin03, McIntosh10}. Within such a picture
we expect that the shortest times (dynamical ruptures) are associated with the 
largest relative
velocities and the longest times are associated with smallest (Coulomb barrier) relative 
velocities.

\begin{figure}
\includegraphics[scale=1.0]{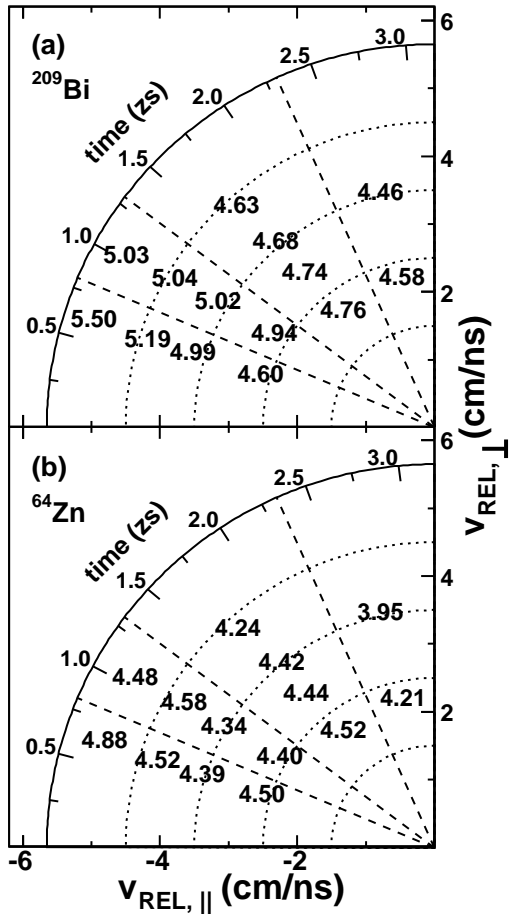}
\caption{\label{fig:Gali_Both} 
Dependence of the average neutron number on the parallel and transverse components
of v$_{REL}$ for Z$_{L}$ = 4 
for the reactions $^{64}$Zn + $^{209}$Bi, $^{64}$Zn.(See text for details) 
}
\end{figure}

In order to explore the dependence of $\langle$N$\rangle$/Z 
on both
cos($\alpha$) and v$_{REL}$, 
we present the dependence of $\langle$N$\rangle$ of Be fragments in 
velocity space in 
Fig.~\ref{fig:Gali_Both}. 
In this figure, the dependence of the average neutron number, $\langle$N$\rangle$,  of Be fragments 
on the transverse (v$_{REL,\perp}$) and parallel (v$_{REL,\parallel}$) 
components of v$_{REL}$ is depicted. 
The parallel and transverse components
of v$_{REL}$ are calculated with respect to the center-of-mass velocity of the 
Z$_L$-Z$_H$ system.
For reference,
relative velocities between 1.5 and 5.5 cm/ns are indicated as dotted circles while
the angular cuts over which the average neutron number was calculated are represented by dashed lines.

For the  $^{209}$Bi (panel a)  target 
a systematic behavior of $\langle$N$\rangle$ of the Be fragment is observed. As one rotates
clockwise in the two dimensional velocity space, i.e. increasing rotation angle $\alpha$, 
the value of $\langle$N$\rangle$ decreases. For the largest v$_{REL}$, the $\langle$N$\rangle$ of the Z$_L$
fragment decreases
from 5.5 to 4.46, a change of $\approx$ 1 neutron as the Z$_H$-Z$_L$ system rotates by a quarter turn.
For the two most backward angle 
bins one observes that 
$\langle$N$\rangle$ decreases with decreasing v$_{REL,\parallel}$ (from 5.5 to 4.6), while for 
larger rotation angles, 
$\langle$N$\rangle$ increases with decreasing v$_{REL,\parallel}$ (from 4.46 to 4.58). 
From this trend,
one would predict that $\langle$N$\rangle$ and hence 
$\langle$N$\rangle$/Z would decrease with increasing v$_{REL}$ for forward emission.
This expectation is confirmed for this system in agreement with previous observation \cite{Hudan12}.
These trends are also observed for the $^{64}$Zn (panel b) target although 
the magnitude of the change is slightly smaller than in the $^{209}$Bi case.

To extract the time dependence of the $\langle$N$\rangle$ of the Z$_L$ fragment, we utilize 
the rotation angle of the Z$_L$-Z$_H$ dinuclear system as a clock such that the rotation 
time, t is given by:
$t=\alpha/\omega$
where $\omega$ is the angular frequency. 
Hence, 
the quantity to be determined is the 
angular frequency which is given by:
$\omega=(J\hbar)/I_{eff}$
where J is the angular momentum and I$_{eff}$ is the moment of inertia for the
dinuclear system. 
The angular momentum of the dinuclear complex is determined by utilizing a simple model
that describes the statistical decay of a rotating system, appropriate for forward emission.
This model was chosen because it provides a simple way to include the effects of thermal excitation
and collective rotation but may have some limitations. 
The magnitude and direction of
the velocity of the PLF$^*$ are determined by sampling the experimental data for forward emission.
The magnitude of the relative velocity vector between Z$_H$ and Z$_L$ is taken from the Viola
systematics \cite{Viola85} with a width provided from the experimental data. The
in-plane and out-of-plane components of v$_{REL}$ are calculated
relative to the plane defined by the PLF$^*$ and beam direction.
The distribution of the out-of-plane emission of Z$_L$ is taken as:
P(sin$\phi$)=Aexp(-$k^2 sin^2\phi$) where $\phi$ is the out-of-plane angle, $k$ 
represents the width of the distribution, and A is a normalization constant. The model
predictions have been filtered by the detector acceptance and compared to
the experimentally measured angular distributions presented in 
Fig.~\ref{fig:cosalpha}. 
Comparison of the measured and predicted distributions for different values of $k$ indicates
that the magnitude of $k$ is $\approx$ 0.5. 
Within the framework of a fissioning nucleus, the parameter $k$ can be related to angular 
momentum \cite{NuclearFission}:
$J^2 = (2k^2I_{eff}T)/\hbar^2$
where T is the temperature and {$ I_{eff}$} is calculated as:
\begin{math}
I_{eff}={\frac{2}{5}}MR^2F_I.
\end{math}
The mass, M, is approximated as:
\begin{math}
M=m_0c^2A_{PLF^*}
\end{math}
where m$_0$c$^2$ is the rest mass of the nucleon and
\begin{math}
A_{PLF^*}=\left(\frac{A}{Z}\right)_{projectile}Z_{PLF^*}.
\end{math}
The effective radius of the dinuclear configuration is given by R$^2$F$_I$ with 
R=r$_0$A$^{1/3}$ and the deviation from a sphere accounted for by F$_I$ \cite{Carjan92}.
The value of the radius constant r$_0$ is taken as 1.2 fm. 
As F$_I$ 
has not been calculated for a system as light as the PLF$^*$ under consideration, we use the
published value for the significantly heavier nucleus, $^{149}$Tb \cite{Carjan92}.  
Assuming a temperature of T=3-5 MeV for the system undergoing binary decay,
we calculate an angular momentum J=6$\pm$1 $\hbar$.

The timescale deduced in this manner is shown in Fig.~\ref{fig:Gali_Both}. It should 
be noted that the timescale deduced (t $\le$ 3 zs) is consistent with previously 
published results \cite{Casini93,Piantelli02,Hudan12}. For reference, the angular velocity calculated for this light
dinuclear complex
is 0.4-0.5 x 10$^{21}$rad/sec.

Having associated the rotation angle with a timescale, it is now possible to observe two timescales
evident in
Fig.~\ref{fig:Gali_Both}. The first observation is that for this system 
even for times as long as 3 zs, i.e. 900 fm/c, the $\langle$N$\rangle$ of the Z$_L$ fragment
is still changing indicating that N/Z equilibration is a slow process. In addition, 
operating on a faster timescale of $\approx$ 1 zs, the v$_{REL}$ dependence observed
for the most backward angles is overpowered by the Coulomb effect that characterizes forward
emission \cite{Hudan05}.
The pattern observed for Z$_L$=4 in 
Fig.~\ref{fig:Gali_Both} indicates that even for backward angles a correlation
exists between the rotation angle dependence and the v$_{REL}$ dependence for
$\langle$N$\rangle$. However, disentangling the intrinsic N/Z gradient from the
Coulomb contribution at backward angles
requires knowledge of the detailed configuration of the dinuclear system. 
The trajectory of the Z$_L$ fragment depends on the motion of the Z$_L$ fragment relative to 
both the target-like and Z$_H$ fragments which
perturbs the intrinsic N/Z pattern. 
This disentanglement is beyond the scope of the present work. We therefore examine the
dependence of $\langle$N$\rangle$ on time, integrated over v$_{REL}$.

\begin{figure}
\includegraphics[scale=0.68]{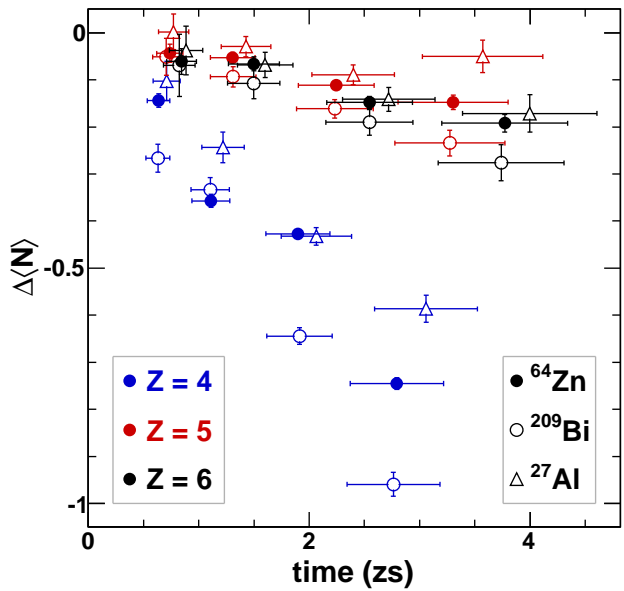}
\caption{\label{fig:dN_time} (Color online) 
Dependence of the average differential neutron number on time
for Z$_{L}$ = 4, 5 and 6 
for the reactions $^{64}$Zn + $^{64}$Zn, $^{209}$Bi and $^{27}$Al.(See text for details) 
}
\end{figure}

Shown in Fig.~\ref{fig:dN_time} is the time dependence of $\langle$N$\rangle$ 
for Z$_L$=4, 5, and 6 fragments. To compare the change in $\langle$N$\rangle$ 
for different targets and different Z$_L$ fragments, we subtracted the extrapolated value
of $\langle$N$\rangle$ at t=0. This extrapolated value was determined by performing
a linear fit of $\langle$N$\rangle$ versus time. It is interesting that the data shown
in Fig.~\ref{fig:dN_time} clearly fall into two groups. While Z$_L$=4 exhibits a strong 
dependence of $\Delta$$\langle$N$\rangle$ on time, Z$_L$=5 and 6 fragments manifest a much weaker
dependence. In the case of Z$_L$=4 fragments, a strong and clear 
target dependence is evident. Close examination of the data for Z$_L$=5 and 6 
fragments reveals that the same target dependence exists though it is smaller 
in magnitude.
For Z$_L$ = 4, the Bi target is associated with the largest change 
in $\langle$N$\rangle$, $\approx$1.
The Zn and Al targets exhibit smaller changes of 0.75 and 0.6 respectively. The horizontal error 
bars shown in the figure are primarily governed by an estimated uncertainty 
of $\Delta$J=1$\hbar$.
It should be noted that the magnitude of the change in 
$\langle$N$\rangle$ is considerably larger
than that previously reported for the Xe + Sn system \cite{Hudan12}.

For Be fragments the large slope of $\Delta$$\langle$N$\rangle$ versus time
indicates significant equilibration occurs on the timescale of 3 zs. One can imagine that
the rate of equilibration is governed by the initial difference in 
N/Z between the Z$_L$ and
Z$_H$ fragments, i.e. a gradient in N/Z within the PLF$^*$. 
Confirming this perspective is the behavior of Be fragments for different targets. 
The data for the Bi target 
manifests the largest slope and the one for the Al target the smallest slope 
corresponding directly to the
neutron-enrichment of the Z$_L$ fragment as evident in 
Fig.~\ref{fig:NZ_CosAlpha}. This target dependence suggests a fundamental result. 
The neutrons transferred from the target do {\it not} equilibrate within the
PLF$^*$ prior to it attaining the dinuclear configuration. If they did, all three targets
would manifest the same equilibration rate.
Thus the experimental data directly indicate that the equilibration rate of
$\langle$N$\rangle$ is governed by the initial N/Z gradient in the 
dinuclear system.

In summary, to investigate the N/Z equilibration timescale for a 
dinuclear complex, 
we have focused on the binary decay of a PLF$^*$. For this system, 
we have shown that the isotopic composition ($\langle$N$\rangle$/Z) 
of the fragments 
depends on the rotation angle of the dinuclear complex. Moreover, the 
N/Z of the
Z$_L$ fragment depends on the neutron-richness of the target. While we 
observe these effects
for Z$_L$=4, 5, and 6 fragments, they are strongest in the case of Be 
fragments. 
This large magnitude for Be fragments can be qualitatively understood as due to the absence 
of $^8$Be in the isotopic distribution making Be a sensitive 
probe of the equilibration process.
Within our analysis, the N/Z equilibration
is related to the change in average neutron number, $\Delta$$\langle$N$\rangle$, experienced
by the Z$_L$ fragment.
Using the 
rotation angle of the dinuclear complex we deduce the timescale on which the N/Z equilibration
occurs. 
The equilibration persists for as long as 3-4 zs. 
Clear evidence for a target
dependence of the equilibration rate is observed in the case of Be fragments. 
The largest equilibration
rate (Bi target) is associated with the largest initial neutron-enrichment of the
Z$_L$ fragment. 
These differences in equilibration rate can be related to different initial N/Z
gradients within the dinuclear configuration for the different targets.
While qualitatively the
physical picture associated with these observations is clear, a more detailed understanding
of the N/Z equilibration will require comparison with a theoretical model that describes
nucleon transport within the dinuclear complex.

\begin{acknowledgments}
This work was supported by the U.S. Department of Energy under 
Grant Nos. DEFG02-88ER-40404 (IU) and DE-FG03-93ER40773 (TAMU). Support 
from the Robert A. Welch Foundation through Grant No. A-1266 is gratefully 
acknowledged.
Collaboration members from Universit{\'e} Laval
recognize the support of the Natural Sciences and Engineering Research Council 
of Canada.
\end{acknowledgments}

\appendix

\bibliography{TAMU_NZ}

\end{document}